\documentclass{article}
\usepackage[utf8]{inputenc}
\usepackage{pdfpages,mathrsfs,amsmath, bm, graphicx, subcaption, amssymb, fancyhdr}
\usepackage[colorlinks,linkcolor=red,anchorcolor=red,urlcolor=black,citecolor=red]{hyperref}
\usepackage[ruled,vlined]{algorithm2e}

\newcommand*{\dif}{\mathop{}\!\mathrm{d}}
\newcommand{\argmin}{\arg\,\min}

\title{Beyond Fourier transform: super-resolving optical coherence tomography}
\author{Yuye Ling$^{1, *}$, Mengyuan Wang$^1$, Yu Gan$^2$, Xinwen Yao$^{3, 4, 5}$, \\Leopold Schmetterer$^{3, 4, 5, 6}$, Chuanqing Zhou$^{7, 8, *}$, Yikai Su$^9$}
\date{\footnotesize{ $^1$ John Hopcroft Center for Computer Science, Shanghai Jiao Tong University, Shanghai, China\\%
    $^2$ Department of Electrical and Computer Engineering, The University of Alabama, AL, USA\\
    $^3$ SERI-NTU Advanced Ocular Engineering (STANCE), Singapore\\
    $^4$ Institute for Health Technologies, Nanyang Technological University, Singapore\\
    $^5$ Singapore Eye Research Institute, Singapore National Eye Centre, Singapore\\
    $^6$ School of Chemical and Biomedical Engineering, Nanyang Technological University, Singapore\\
    $^7$ Institute of Biomedical Engineering,  Shenzhen Bay Lab, Shenzhen, China\\
    $^8$ School of Biomedical Engineering, Shanghai Jiao Tong University, Shanghai, China\\
    $^9$ State Key Lab of Advanced Optical Communication Systems and Networks, Department of Electronic Engineering, Shanghai Jiao Tong University, Shanghai, China \\
    $^*$\href{mailto:yuye.ling@sjtu.edu.cn}{yuye.ling@sjtu.edu.cn} and \href{mailto:zhoucq@szbl.ac.cn}{zhoucq@szbl.ac.cn}\\[2ex]%
    \today}}

\begin{document}
\maketitle

\begin{abstract}
Optical coherence tomography (OCT) is a volumetric imaging modality that empowers clinicians and scientists to noninvasively visualize the cross-sections of biological samples. As the latest generation of its kind, Fourier-domain OCT (FD-OCT) offers a micrometer-scale axial resolution by taking advantage of coherence gating. Based on the current theory, it is believed the only way to obtain a higher-axial-resolution OCT image is to physically extend the system’s spectral bandwidth given a certain central wavelength. Here, we showed the belief is wrong. We proposed a novel reconstruction framework, which integrates prior knowledge and exploits the \emph{shift-variance}, to retrospectively super-resolve OCT images without altering the system configurations. Both numerical and experimental results confirmed the processed image manifested an axial resolution beyond the previous theoretical prediction. We believe this result not only opens new horizons for future research directions in OCT reconstruction but also promises an immediate upgrade to tens of thousands of legacy OCT units currently deployed.
\end{abstract}

The everlasting technological advancement in optical imaging is one of the driving forces that are transforming the landscape of modern healthcare and medical research \cite{Huang1991, Hell1994, Greenbaum2012, Bouchard2015}. The success of OCT in ophthalmology is a prominent example: in less than 30 years, this volumetric imaging technique has evolved from a bench-top research tool into one of the gold standards in clinics. It is estimated now that more than 30 million OCT exams are performed annually across the globe \cite{Swanson:17}. There is therefore a continuing interest in elevating the performance of OCT, especially for its axial resolution.

Fourier-domain OCT (FD-OCT), the latest generation of OCT, typically possesses an axial resolution ranging from a few to tens of micrometers. Unlike its time-domain counterpart, FD-OCT utilizes a simple inverse discrete Fourier transform (IDFT) to recover the axial profile of the object from its spectral interferogram without performing physical axial scans \cite{RN670}, which has enabled numerous exciting applications including real-time imaging \cite{Wieser2014} and functional imaging \cite{Kennedy2017, Boer2017}. Based on the current FD-OCT theory, it is believed that the axial resolution of the reconstruction is theoretically and ``physically'' limited by the full-width-half-maximum (FWHM) of the light source’s coherence function, which is inversely proportional to its spectral bandwidth \cite{mandel_wolf_1995, Izatt2015}.

Although various signal processing techniques including deconvolution \cite{612153}, spectral shaping \cite{Tripathi:02}, and numerical dispersion compensation \cite{Fercher2001, Marks2003} have been widely deployed in FD-OCT, they are mostly designed to improve image quality instead of raising image resolution. Novel techniques such as sparse representation \cite{Abbasi} and deep learning \cite{Halupka:18, Huang:19} are starting to gain traction in retrieving high-resolution images from low-resolution ones, despite requiring a learning or training process. Nonetheless, the axial resolution of the post-processed images would still be ultimately capped by the aforementioned theoretical limit, because the high-resolution images used for training were still obtained by using FFT technique.

As a result, ``improving the axial resolution of an FD-OCT system'' has become synonymous with ``increasing emission bandwidth'' in the community \cite{Cense2004}: research groups tend to solely measure the FWHM of the system's axial point spread function (PSF), and report it as the system's axial resolution \cite{RN141}. To date, multiple groups have showcased FD-OCT with about 1-$\mu$m axial resolution according to this convention, and most of them are achieved by using an 800-nm system with an extra-large bandwidth of about 300 nm \cite{Liu2011, Yadav2011, Bizheva2017}. It is worth noting that pushing the axial resolution beyond this mark through a further expansion on the emission bandwidth could be hellacious: this will not only require a technological progression in laser sources but also a proper handling of the collateral chromatic aberration and dispersion.

Here, we argue that both the axial resolution limit and the reciprocity relationship between the axial resolution and the spectral bandwidth are not inherent in OCT but are induced by the IDFT-based reconstruction algorithm. Specifically, we believe the spectral measurements from the FD-OCT does carry ultrastructural information: the FD-OCT system is a coherent imaging system in the axial direction and should not be fully characterized by its axial PSF as its incoherent counterparts \cite{RN69}. To address this issue, we adopted a novel mixed-signal system model and showed that the FD-OCT system is rather \emph{shift-variant} than \emph{shift-invariant}, which contradicts the common belief. We then proposed an optimization-based reconstruction framework to exploit the newly discovered \emph{shift-variance}. In contrast to the IDFT-based reconstruction, we could integrate apparent physical insights of the object, including sparsity, to the solutions within the proposed framework. An axial resolution beyond the previous prediction could thus be achieved. Our claim is supported by simulation, phantom experiments, and biological-relevant experiments. 

Since the proposed method is computational and is developed upon ordinary FD-OCT measurement, it could be retrospectively applied to any existing dataset to improve the axial resolution as long as the raw spectral measurements are accessible. To facilitate this process, we have also made our source codes fully available online.

\section*{Results}
\subsection*{Established FD-OCT theory and its limitations}
First conceived by Fercher \emph{et al} in 1995 under the name of ``backscattering spectral interferometry'' \cite{RN670}, FD-OCT offers a one-dimensional approximate solution to the optical inverse scattering problem \cite{RN650}. The authors derived that the backscattered field amplitude $A_S(k)$ is proportional to the Fourier transform of the scattering potential $F_S(z)$, when the measurement is taken in the far-field from a weakly scattering object \cite{RN670}. This important finding laid the theoretical foundation of FD-OCT: the depth profile of an unknown object could be retrieved by performing an inverse Fourier transform on the spectrum of the backscattered light. However, in most real-world implementations, backscattered light intensity $I(k) = |A_S(k)|^2$ is only available in its truncated \emph{and} sampled form. Unfortunately, most researchers simply use the inverse discrete Fourier transform (IDFT) in place of the inverse Fourier transform to recover the object function without further justifications \cite{RN300, RN298,RN330}. There are at least two issues associated with this practice.


The first issue is the axial resolution. Since the digitization process is \emph{shift-variant}, the entire FD-OCT system would become \emph{shift-variant} as well \cite{easton2010fourier}. Therefore, it is no longer suitable to use the linear \emph{shift-invariant } theory to analyze the FD-OCT system; the axial resolution of the FD-OCT should not be quantified and is not ``theoretically'' limited by the FWHM of the light source's coherence function (or emission spectrum). Additionally, if we consider that FD-OCT is indeed a \emph{coherent} imaging modality, we are entitled to access extra information about the sample location if the phase data is fully exploited \cite{RN69}. Phase-resolved techniques are good examples \cite{RN298, Zhao:00}, by which picometer-scale vibrations were successfully detected \cite{Lee3128}. 

The second issue is the accuracy. The IDFT-reconstructed object is not exact due to the erroneous signal processing: the Fourier transform relationship existed between $F_S(z)$ and $A_S(k)$ in the analog domain does not guarantee the discrete Fourier transform relationship between their sampled counterparts in the digital domain. Therefore, the IDFT-reconstructed object would differ from the original object by digitization errors, alias, and leakage \cite{Briggs}.

\subsection*{Super-resolving OCT: an \emph{ill-posed} inverse problem}
\begin{figure}[htbp]
    \centering
    \includegraphics[width=\textwidth]{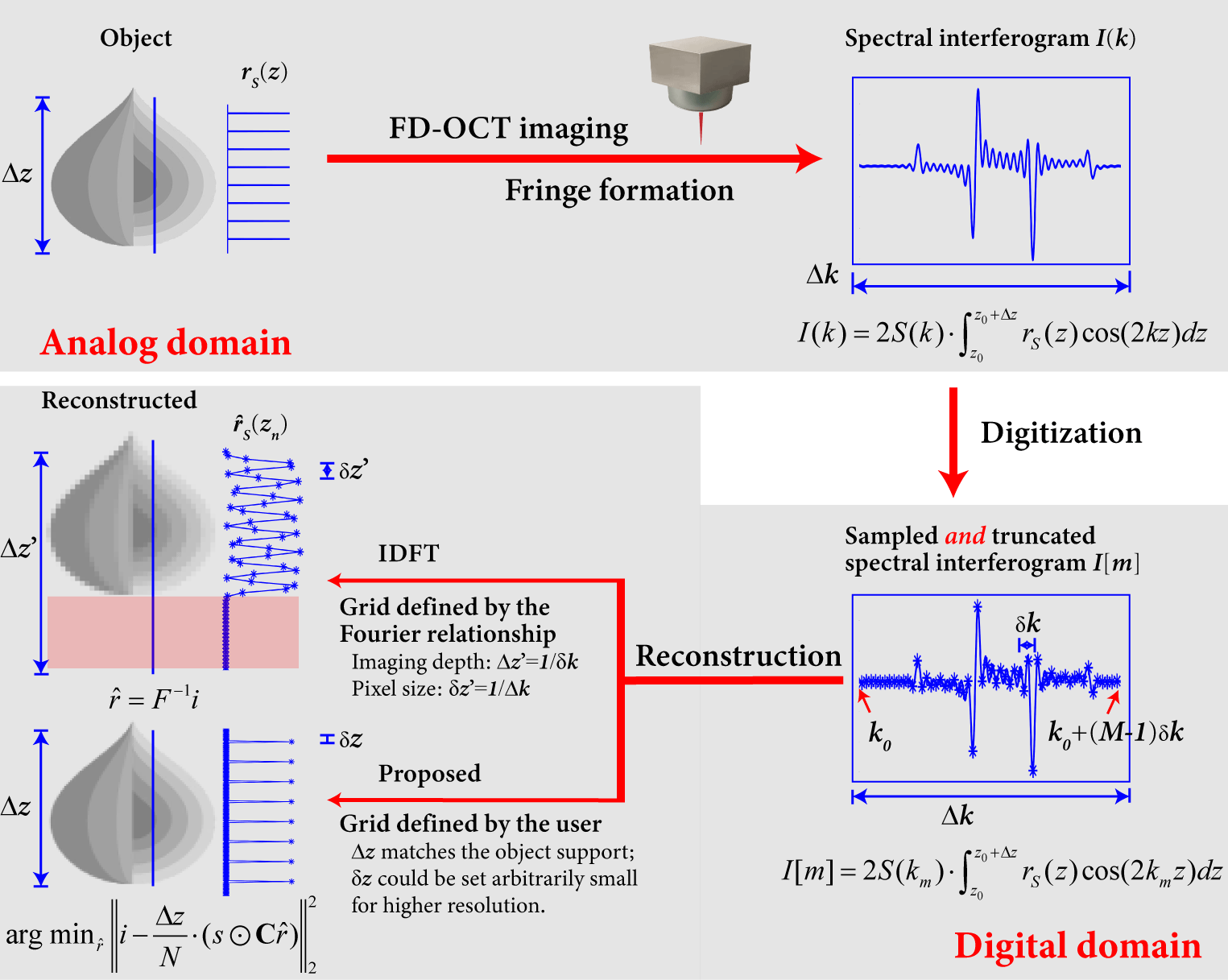}
    \caption{Schematic illustration of the fringe formation and the image reconstruction in FD-OCT. After FD-OCT imaging, the analog object $r_S(z)$ forms an analog spectral interferogram pattern $I(k)$, which will later be digitized by measurement devices, giving samples $I[m]$. Conventionally, an IDFT is performed directly on the resultant digital signal $I[m]$ to reconstruct the object, where the reconstruction resolution $\delta'_z$ and image depth $\Delta z'$ are implicitly defined by the spectral sampling interval $\delta_k$ and bandwidth $\Delta k$. Here we proposed an alternative optimization-based approach to this process: both the grid size $\delta_z$ and image depth $\Delta z$ could be arbitrarily defined by the user to achieve best reconstruction performance.} 
    \label{fig:schematic}
\end{figure}

To address the difficulties existed in the current OCT theory, we proposed a new framework, where the FD-OCT is now modeled as a mixed-signal system as illustrated in Fig. \ref{fig:schematic}: the physical formation of the spectral interferogram $I(k)$ is in the analog domain, while the image reconstruction is in the digital domain. Two domains are interfaced by measurement devices, which perform the analog-to-digital conversion (ADC). The measured spectral interferogram $I[m]$ is given by (Supplementary Section 1),

\begin{equation}
        I[m] = 2 S(k_m) \cdot \int_{z_0}^{z_0 + \Delta z}{r_S(z) \cos{(2 k_m z)} \dif z} + N(k_m), \quad m = 0, \ldots, M - 1,
    \label{Equation1}
\end{equation}
where $S(k_m)$ is the sampled power emission spectrum of the light source, $N(k_m)$ is the noise presented, $k_m$ is the $m^{\textrm{th}}$ sample in the wavenumber domain, $r_S(z)$ is the reflectivity profile (object function) of the sample, and $[z_0, z_0+\Delta z]$ defines its support. 

It is intriguing to observe that the measured spectral interferogram $I[m]$ contains all the information about the analog object function $r_S(z)$. Ideally, by solving the integral equations we should be able to \emph{exactly} recover the $r_S(z)$. Unfortunately, Eq. (\ref{Equation1}) has no closed-form solutions. To numerically solve Eq. (\ref{Equation1}), we discretize its right-hand side uniformly by $N$ times, and rewrite the equation in short form by using matrix-vector notation \cite{RN120} (Supplementary Section 2),
\begin{equation}
    \boldsymbol{i} = \delta_z \cdot (\boldsymbol{s} \odot \mathbf{C} \boldsymbol{r}) + \boldsymbol{n}
    \label{Equation2}
\end{equation}
where $\boldsymbol{i}=\{I[0], I[1], ..., I[M - 1]\}^\intercal$, $\boldsymbol{r}=\{r_S(z_0), r_S(z_1), ..., r_S(z_{N - 1})\}^\intercal$, $\boldsymbol{s}=\{S(k_0), S(k_1), ..., S(k_{M-1}))\}^\intercal$,  $\delta_z = \Delta z/ N$, $\mathbf{C}$ is an $M \times N$ matrix with $C_{mn} = \cos{(2k_mz_n)}$, $\boldsymbol{n}$ denotes the noises, and $\odot$ represents for Hadamard product or element-wise multiplication.

Clearly, the axial resolving capability of the system is now dependant on (1) how well we can solve the Eq. (\ref{Equation2}), and (2) how good Eq. (\ref{Equation2}) can be used to approximate Eq. (\ref{Equation1}) numerically. If we undersample the object function $r_S(z)$ by using a relatively small $N$ ($N \leq M$), we can solve Eq. (\ref{Equation2}) with ease at the cost of having a large approximation error. This is the case we have in conventional IDFT-based reconstruction algorithm (Supplementary Section 3). 

On the other hand, we could also choose to oversample $r_S(z)$ by using a very large $N$ ($N \gg M$). In this case, the approximation error will be minimized and the axial resolution of the reconstruction will only be limited by the sampling interval. But solving Eq. (\ref{Equation2}) in this case is nontrivial, since it is now an \emph{ill-posed} inverse problem \cite{bertero1998introduction}: the aim is to find $M$ unknowns ($N$-dimensional high-resolution object $\boldsymbol{r}$) from $M$ equations ($M$-dimensional measurement $\boldsymbol{i}$).

\subsection*{Proposed reconstruction via constrained optimization}

Fortunately, the OCT image reconstruction is a physical problem: the solution that we are seeking should not only fit the mathematical equations described above but also satisfy some additional physical constraints. We can, therefore, apply regularizations to Eq. (\ref{Equation2}) to acquire a feasible solution even if it is \emph{ill-posed}.

One such \emph{prior} information available is the predefined support for the object $\boldsymbol{r}$ \cite{RN847}. For example, a rough estimation of the support could be obtained from the low-resolution $\hat{\boldsymbol{r}}_\textrm{IDFT}$ by thresholding.

Another \emph{a prior} we could utilize is the \emph{sparsity} of the object $\boldsymbol{r}$. Considering that $N$ is large and $\delta_z$ is only a fraction of the incident wavelength, it is plausible for us to assume the refractive index distribution of the object to be slow varying and piecewise constant at this scale. This will thus make the object's reflectivity profile $\boldsymbol{r}$, which is the first derivative of the refractive index distribution, to be \emph{sparse} \cite{Mousavi2016}. Therefore, we could use $\ell1$-norm minimization to promote the \emph{sparsity} in the solution \cite{RN687, RN688}. It is worth noting that the \emph{sparsity} presented here is independent of the object's structure, but rather relies on the sampling density. 

We would like to focus on using the latter constraint in the current study, although there are infinitely many possible assumptions we could reasonably make towards the structure of the solution \cite{LUBK201859}. The OCT reconstruction can thus be formatted as a constrained optimization problem, 
\begin{equation}
    \begin{array}{ll}
     \text{minimize}   &  \lVert \hat{\boldsymbol{r}} \rVert_1 \\
     \text{subject to}  & \lVert \boldsymbol{i} -  \delta_z \cdot (\boldsymbol{s} \odot \mathbf{C} \boldsymbol{r})\Vert_2^2 \leq \epsilon
    \end{array}
    \label{Equation4}
\end{equation}
where $\hat{\boldsymbol{r}}$ is the reconstructed object discretized on a grid that is \emph{arbitrarily} defined by the user. 

\subsection*{Axial \emph{super-resolution} demonstration}
\begin{figure}[htbp]
    \includegraphics[width=\textwidth]{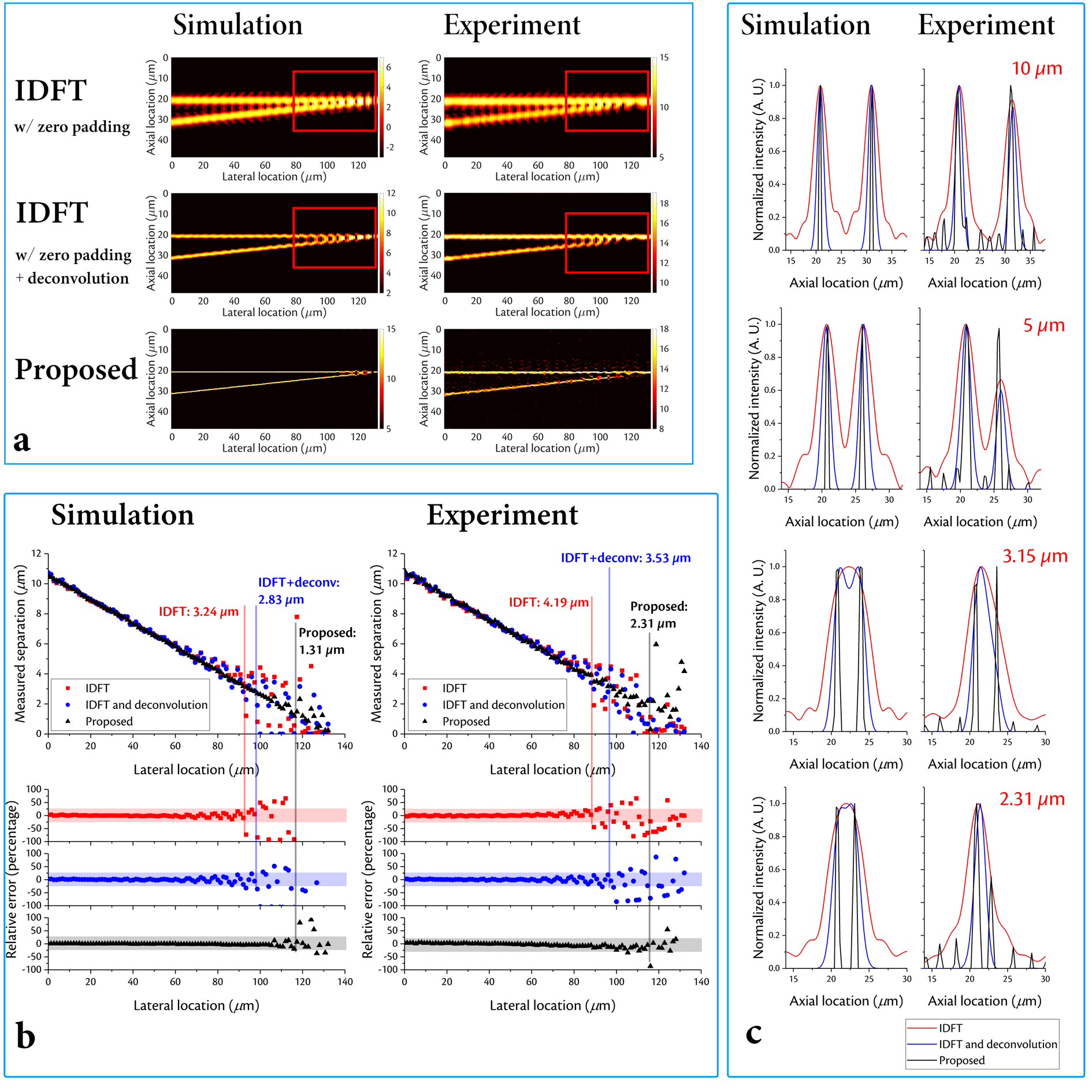}
    \caption{Axial \emph{super-resolution} capability. Three reconstruction techniques including direct IDFT, direct IDFT followed by Lucy-Richardson deconvolution \cite{RN144}, and the proposed technique are used to reconstruct an air wedge. The corresponding cross-sectional images are shown in (a). Speckle patterns are observed in both IDFT-based reconstructions as pointed out by the red boxes. The axial location of the upper and the lower interface of the air wedge are fitted by using a two-term Gaussian model. The separation is calculated and plotted in the top row of (b). The relative error between the measured separation and the predicted value is plotted in the bottom row of (b). The axial resolution of a given technique is determined when the deviation is greater than 20\%: the axial resolution of the proposed system is 2.31 $\mu$m, which is below the theoretical value. To further illustrate the idea, the A-line profiles at 4 different separations (10 $\mu$m, 5 $\mu$m, 3.15 $\mu$m, 2.31 $\mu$m) are given in (c). The proposed technique clearly outperforms its IDFT-based counterparts. Simulation results for all the experiments are also provided for comparison.}
    \label{fig:air_wedge}
\end{figure}

To characterize the axial resolution of various reconstruction techniques, we constructed an air wedge and imaged it by using a commercial FD-OCT system. The results obtained by using direct IDFT, direct IDFT followed by Lucy-Richardson deconvolution \cite{Liu:09}, and the proposed optimization-based algorithm are presented in Fig. \ref{fig:air_wedge}a. By visual inspection, the proposed technique offers not only the highest axial discernment capability but also a superior image quality with suppressed speckles (or ``interference fringe pattern'' \cite{Liu2015}) which are caused by the coherent addition of the backscattered waves from the air wedge's interfaces.

We then measured the separation between the interfaces from the reconstructed images, and the results are plotted in Fig. \ref{fig:air_wedge}b. The measured axial resolution by using the proposed technique is $\sim$2.31 $\mu$m, which is well below the FWHM of the coherence function ($\sim$3.40 $\mu$m, See Methods). For reference, the measured axial resolution by using direct IDFT and direct IDFT followed by deconvolution is 4.19 $\mu$m and 3.53 $\mu$m, respectively. Both are limited by the FWHM of the coherence function.

After that, we plotted A-line profiles obtained by aforementioned techniques for selected separation (10 $\mu$m, 5 $\mu$m, 3.15 $\mu$m, and 2.31 $\mu$m) in Fig. \ref{fig:air_wedge}c. As expected, both the 10-$\mu$m and 5-$\mu$m separation could be discriminated by all three methods. However, only the proposed optimization-based reconstruction is able to resolve the 3.15-$\mu$m and 2.31-$\mu$m separation. 

Finally, we repeated the experiments numerically (See Methods), and the simulation results are placed side by side with the corresponding experimental results in Fig. \ref{fig:air_wedge}. The experimental results bears a very good resemblance to the numerically synthesized ones despite of a slight degradation in performance. It is worth noting that the simulation is noise-free, while the experiment is not.

\begin{figure}[htbp]
        \includegraphics[width=\textwidth]{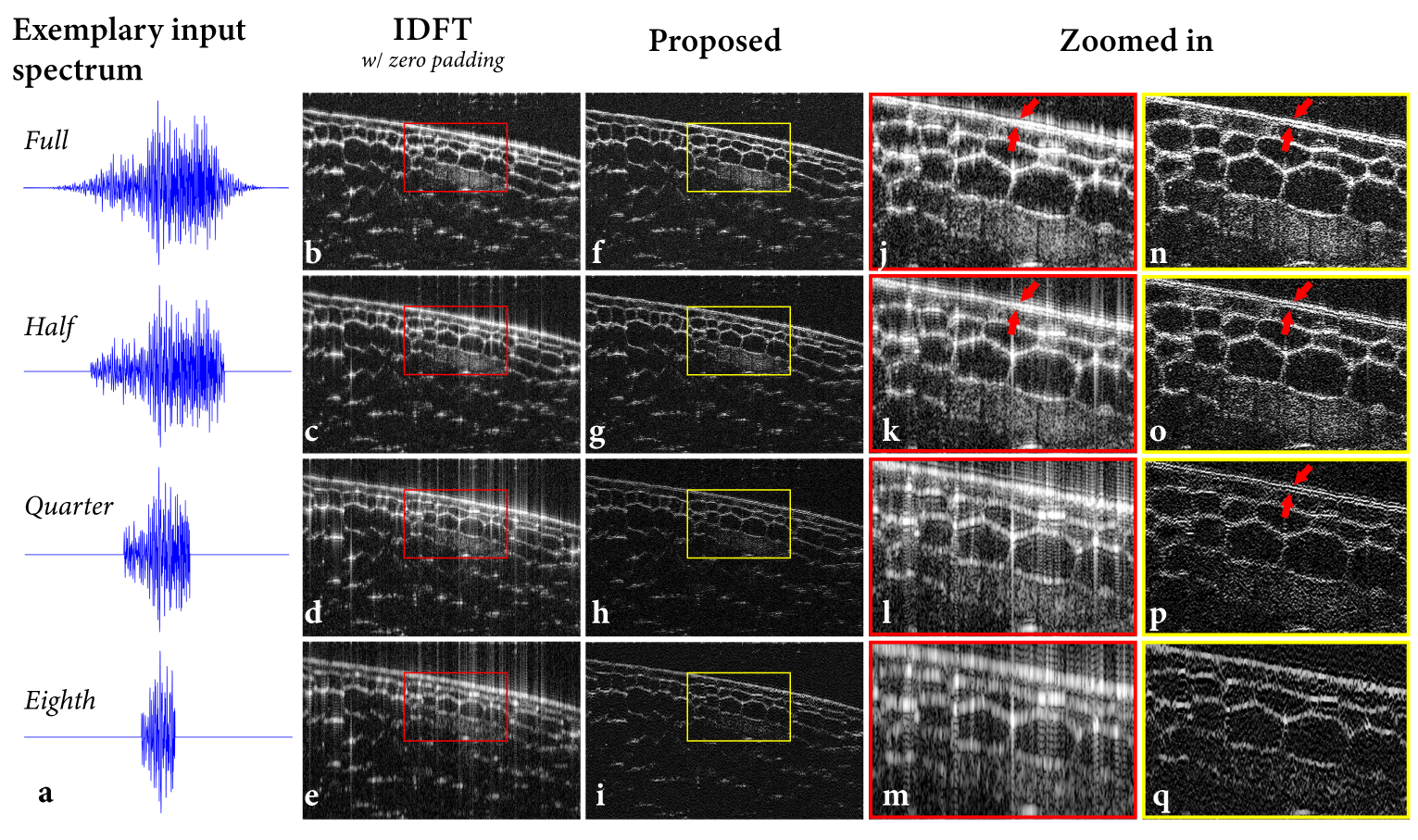}
        \caption{Single-frame comparisons of the reconstructed onion obtained by IDFT and the proposed technique. (a) Exemplary input spectrum for different spectral bandwidth. Different bandwidths are obtained via numerical truncation. (b)-(e), Reconstructed images obtained by using conventional IDFT from full bandwidth, half bandwidth, quarter bandwidth, and eighth bandwidth, respectively. (f)-(i), Reconstructed images obtained by using optimization-based technique from the same input as in (b)-(e). (j)-(m), Magnified views of the regions highlighted in (b)-(e). (n)-(q), Magnified views of the regions highlighted in (f)-(i). The red arrows point out the double-layer structure of the onion skin that could be visually distinguished.}
    \label{fig:onion}
\end{figure}

\begin{figure}[htbp]
        \includegraphics[width=\textwidth]{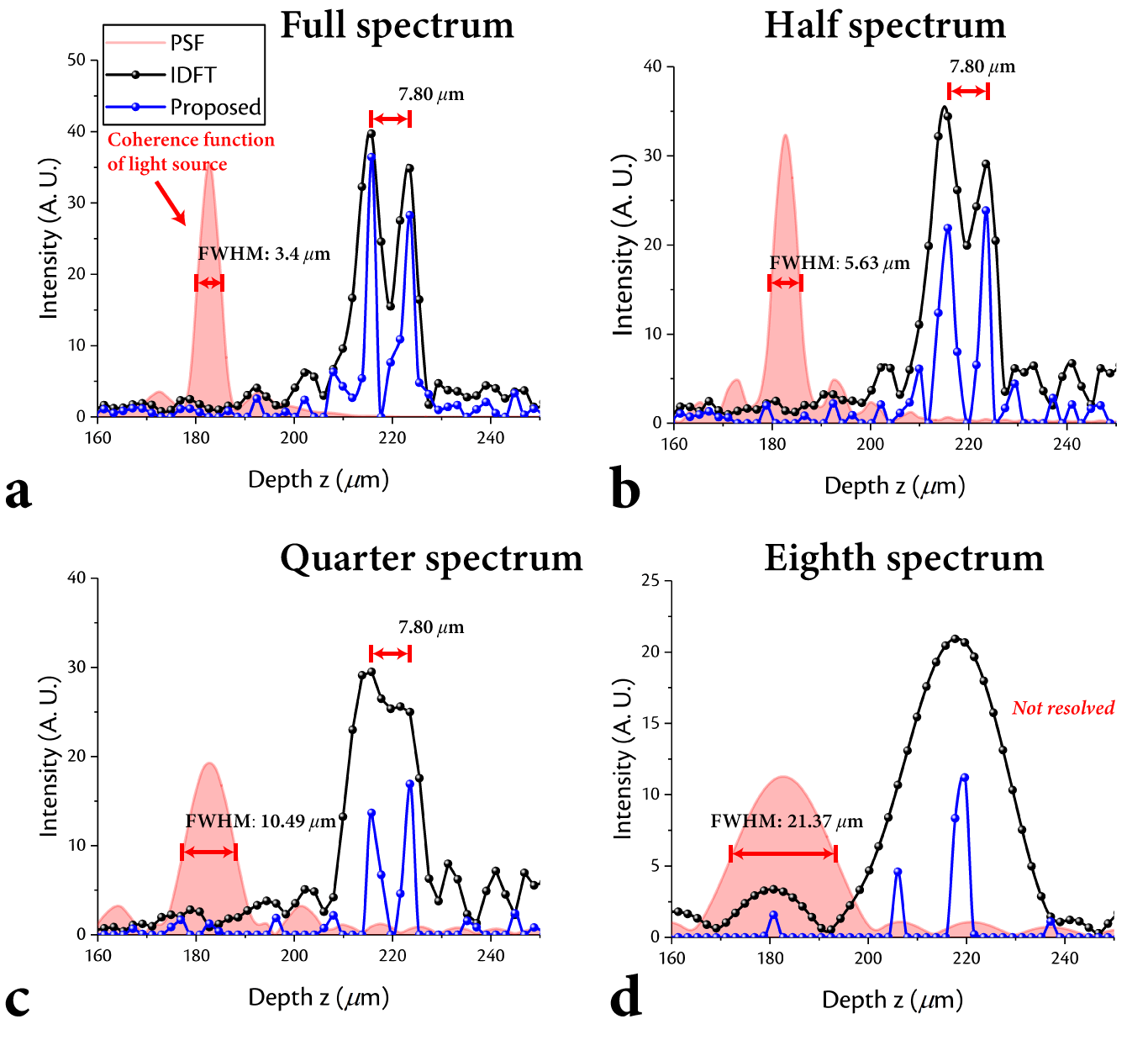}
        \caption{The super-resolving capability of the proposed technique. A lateral position within the region pointed out by red arrows in Fig. \ref{fig:onion} is selected. The A-line profile excerpts at the selected location reconstructed by using both IDFT and proposed techniques from (a) full bandwidth, (b) half bandwidth, (c) quarter bandwidth, and (d) eighth bandwidth are plotted, respectively. The corresponding theoretically predicted point spread functions and their FWHM are also given for reference. The separation between the two layers of the onion skin is measured 7.8 $\mu$m. The proposed method managed to resolve this structure even when the theoretical resolution is larger (c.f. (c)).}
    \label{fig:aline}
\end{figure}

\subsubsection*{\emph{Super-resolving} biological samples}

To further validate our method, we applied our technique on onion images acquired from the same commercial FD-OCT. We first reconstructed images from full-spectrum measurements, which is used as references as shown in Fig. \ref{fig:onion}b-e. We then repeated the procedures for reduced-bandwidth measurements as illustrated in Fig. \ref{fig:onion}a, in which we mimicked the scenarios of having lower-resolution systems. We compared the images obtained from the latter (Fig. \ref{fig:onion}f-i and n-q) with the references. Specifically, we visually examined whether the double-layer skin structure pointed out by the red arrows is visible in various reduced-bandwidth cases. By using the proposed method, we could even distinguish the double-layer structure when the bandwidth is quartered (Fig. \ref{fig:onion}m). Finally, we plotted the A-line profiles of the onion skin obtained by both methods for different bandwidth cases in Fig. \ref{fig:aline}. The theoretically predicted point spread functions for each case are also provided for references. It is confirmed that the proposed technique does enable higher resolution in experimental situations and outperforms its IDFT-based counterpart as illustrated in Fig. \ref{fig:aline}c. In fact, the proposed method achieved a resolution higher than the theoretical value just as predicted in last section.

\begin{figure}
    \centering
    \includegraphics[width = 0.9\textwidth]{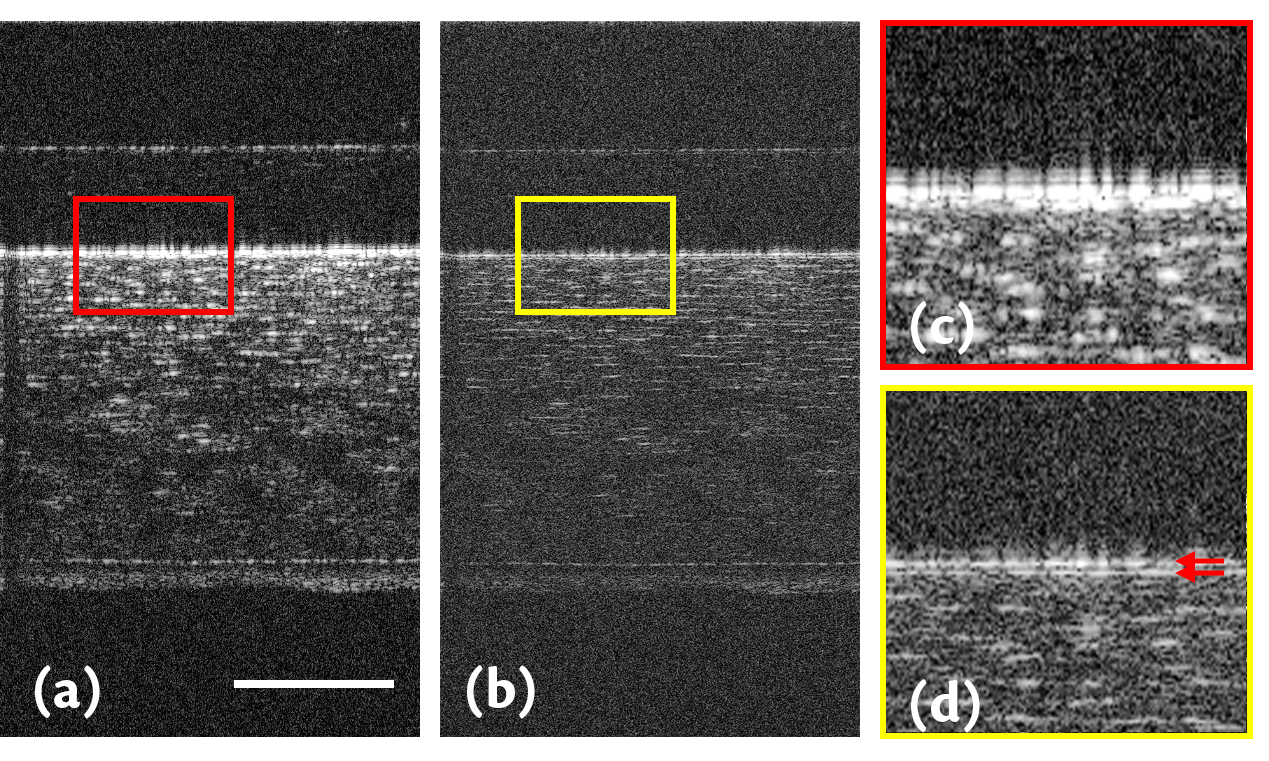}
    \caption{\emph{In vivo} monkey cornea reconstructed by (a) IDFT and (b) the proposed technique. Magnified views of the regions highlighted in (a) and (b) are shown in (c) and (d), respectively. The proposed methods shows better contrast and could clearly delineate the Descemet’s membrane as pointed out by the red arrow.}
    \label{fig:monkey}
\end{figure}

We also applied our technique on an existing dataset to retrospectively \emph{super-resolve} the images. Specifically, the original data was obtained from a custom-built ultra-high-resolution SD-OCT \cite{Werkmeister:17} on \emph{in vivo} monkey cornea \cite{Yao:19}. The conventional IDFT-based reconstruction is presented in Fig. \ref{fig:monkey}a, which is the same one rendered as Fig. 5a in \cite{Yao:19}, and its \emph{super-resolved} counterpart is plotted in Fig. \ref{fig:monkey}b. The super-resolved image shows better axial resolution with slightly reduced signal-to-noise ratio (SNR). Moreover, the Descemet's membrane (has a typical thickness of 5-10 $\mu$m), which could barely be resolved, is clearly observed in Fig. \ref{fig:monkey}b. Magnified views of the same region are illustrated in Fig. \ref{fig:monkey}c and d for better comparison.

\section*{Discussion}

Despite the wide-spread popularity of FD-OCT as a volumetric imaging technique, image reconstruction from spectral interferogram has largely relied on simple IDFT. To enhance quality of reconstructed images, a large body of literature has explored signal processing techniques for post-processing. However, unlike in other branched of biomedical imaging, research on FD-OCT rarely touch upon the reconstruction procedure itself. In the current work, we filled this gap by framing reconstruction as an inverse problem and demonstrated great potential of using such framework on FD-OCT.

Although it is not analytically proved, we numerically showed that the axial resolution of the proposed technique is mostly SNR-limited (Supplementary Section 4), which would be inline with most optimization-based image reconstruction algorithms \cite{RN41}. In addition to that, the axial resolution also depends on the grid step as well as the accuracy of the physical assumptions we have made towards the original object (e.g. in this study, whether the object is indeed \emph{sparse} at a given sampling density).

Currently, the processing speed of the proposed algorithm is slow for \emph{in vivo} applications. But in the future, we would like to explore the deep learning approach to solve the inverse problem, which would promise a significant speed-up to the entire processing \cite{Adler_2017}. 

In conclusion, a novel image reconstruction algorithm for FD-OCT based on optimization is presented. By exploiting the physical priors of the object, the proposed technique can outperform its IDFT-based counterpart, and the result is validated both numerically and experimentally. 

A key theoretical contribution of this piece of work is the recolonization of the importance of digitization in both FD-OCT imaging and reconstruction process: (1) the sampling during the spectral measurement compromises the elegant \emph{shift-invariance} of the system and (2) a user-defined arbitrary sampling grid in reconstruction domain could help break the axial resolution limit imposed by the conventional IDFT-determined grid. This work shall offer a new perspective to the FD-OCT community as to the image reconstruction problem: IDFT-based construction might not be the only solution not to mention its optimality. 

In the future, more sophisticated physical priors could be integrated to the proposed framework to further improve the image quality. Theoretical analysis on the convergence of the algorithm, the lower bound of the resolution, as well as the robustness of the algorithm would also be the future direction of this study. We believe the proposed the technique to be extremely valuable considering that it is a pure algorithmic efforts: an FD-OCT could get a boost in axial resolution without any hardware modifications.

\section*{Acknowledgement}
The authors would like to thank Prof. Bo Jiang, Dr. Wenxuan Liang and Mr. Tingkai Liu for their enlightening discussions. This project is supported by National Natural Science Foundation of China (NSFC) (61905141, 61875123), Shanghai Science and Technology Committee (1744190202900), and Shanghai Jiao Tong University Interdisciplinary Research Fund of Biomedical Engineering (YG2016ZD07). We acknowledge the computing resources from Shanghai Jiao Tong University high-performance computing facilities $\pi$2.0-cluster.

\bibliographystyle{naturemag}
\bibliography{main}

\section*{Methods}
\subsection*{Proposed optimization-based reconstruction algorithm}
Although the proposed technique can incorporate various physical constraints into the reconstructed image, we focus on the use of sparsity properties of the sample here as an example.

To mathematically solve this problem, Eq. (\ref{Equation4}) is reformatted to its Lagrange form,

\begin{equation}
    \argmin_{\hat{\boldsymbol{r}}} \frac{1}{2} ||i-\frac{\Delta z}{N} \cdot \mathbf{C} \hat{\boldsymbol{r}}||_2^2 + \lambda||\hat{\boldsymbol{r}}||_1
    \label{Equation5}
\end{equation}
where $\lambda$ is the Lagrange multiplier. We employed the Alternating Direction Method of Multipliers (ADMM) \cite{MAL-016} to solve Eq. (\ref{Equation5}), and the derived algorithm is summarized in Algorithm 1.
 
\begin{algorithm}[h]
\caption{ADMM algorithm for solving Eq. (\ref{Equation4})}
\begin{align*}
  \textbf{Problem:} & \ \argmin_{\hat{\boldsymbol{r}}} \frac{1}{2} ||i-\frac{\Delta z}{N} \cdot \mathbf{C} \hat{\boldsymbol{r}}||_2^2 + \lambda||s||_1 \quad \textrm{s.t.} \quad \hat{\boldsymbol{r}}-s=0 & \\
  \textbf{Input:}   & \ \mathbf{C} \; \textrm{in} \; R^{m \times n},i\; \textrm{and} \; \rho, \lambda, \alpha\\
  \textbf{Initialize:} & \ \hat{\boldsymbol{r}}^{(0)}, s^{(0))}, u^{(0)}\ \textrm{where}\  u^{(0)}=\frac{1}{\rho}y^{(0)}\\
  \textbf{While:} & \ \hat{\boldsymbol{r}}^{(k)}, s^{(k))}, u^{(k)}\ \textrm{do not converge (or reach maximum iteration)}\\
    \hat{\boldsymbol{r}}^{(k+1)}& \ \leftarrow{\argmin}_{\hat{\boldsymbol{r}}}\textbf{L}(\hat{\boldsymbol{r}}^{(k)},s^{(k))}, u^{(k)})\\
                    & \ =\argmin \ f(\hat{\boldsymbol{r}}^{(k)})+\frac{\rho}{2}||\hat{\boldsymbol{r}}^{(k+1)}-v^{(k)}||_{2}^{2}\\
                    & \ =(\mathbf{C} \mathbf{C}^{T} + \rho )^{-1} (\mathbf{C}^{-1} b + \rho v^{(k)})\\
    s^{(k+1)} & \ \leftarrow  {\argmin}_{s} \textbf{L}(\hat{\boldsymbol{r}}^{(k+1)}, s^{(k))}, u^{(k)})\\
                    & \ =\argmin \ f(\hat{\boldsymbol{r}}^{(k+1)})+\frac{\rho}{2}||\hat{\boldsymbol{r}}^{(k+1)}-v^{(k)}||_{2}^{2}\\
     u^{(k+1)}& \ \leftarrow {u^{(k)}} + \hat{\boldsymbol{r}}^{(k+1)}-s^{(k+1)}\\
\textbf{end While}&\\
\textbf{Output:}& \ \hat{\boldsymbol{r}}^{*} \longleftarrow \hat{\boldsymbol{r}}
\end{align*} 
\end{algorithm}
The algorithm is implemented in GNU Octave (version 4.4.1) on a high-performance computing facility (Shanghai Jiao Tong University $\pi2.0$-cluster, China).  It takes about 4s to reconstruct an A-line with 2,048 sampling points. The maximum iteration number is set to 1,000 to balance the reconstruction speed and the quality.

\subsection*{Experimental set-up}
Both the air wedge and the onion imaging were performed on a commercially available high-resolution spectral-domain optical coherence tomography (SD-OCT) system (GAN620C1, Thorlabs, USA). The system is centered at 892.8 nm, and the spectrometer measures from 791.6 nm to 994.0 nm with 2,048 pixels. The reconstruction grid step and imaging depth, based on IDFT method, are 1.94 $\mu$m and 1.9 mm, respectively. The grid size and support size for the optimization-based method was set to 1 $\mu$m and 1 mm, respectively.

We extracted the $k$-linearized spectra directly from the SD-OCT system. No preprocessing was performed. To obtain the reduced-bandwidth measurement, we numerically truncated the original measurement as illustrated in Fig. \ref{fig:onion}a. The corresponding theoretical PSF is obtained by directly performing IDFT on the truncated emission spectra. During the reconstruction, the Lagrange multiplier $\lambda$ for full, half, quarter, and eighth bandwidth is 1,000, 1,000, 100, and 10, respectively. 

\subsection*{Details about the air wedge experiment}

The air wedge was constructed by placing one microscope cover glass on top of the other, while a segment of tape is inserted on one edge. During the B-scan imaging, the entire lateral field-of-view was set to be 4 mm, which is significantly larger than the cross-sectional image given in Fig. \ref{fig:air_wedge}.

To extract the axial location of the top and bottom interface of the air wedge, we used a two-term Gaussian model to fit the A-line profile. We then recorded the centers for both interfaces, subtracted one from the other, took the absolute value, saved it as the measured separation for a given lateral position, and plotted it in the top row of Fig. \ref{fig:air_wedge}b.

Based on the assumption that the cover glass is a rigid body with straight surface, the measured separation should be linear against lateral position. We then linearly fitted the data and obtained the expected separation at every lateral position. It is worth noting that the coefficient of determination ($R^2$) is 1.

We then calculated the relative error between the measured separation v.s. the expected separation and plotted it in the bottom row of Fig. \ref{fig:air_wedge}b. It is clear that the relative error for a given technique would start to diverge beyond a certain axial separation. Here, we empirically defined the axial resolution to be the minimal separation that the relative error is smaller than 20\%.

Since we already obtained the expected separation via linear fitting, we used it as ground truth to numerically synthesize the simulation results by using a piece of OCTAVE codes.





\end{document}


\maketitle

\section{An inconvenient truth about the digitization in FD-CT}

In the following derivations, we will use power reflectivity and scattering potential interchangeably.

Assume the weakly scattering object under investigation has a reflectivity profile $r_S(z)$, which is a real-valued function defined on a finite support of $[z_0, z_0 + \Delta z]$ as illustrated in main text Fig. (1). The scattered electric field $E_S(k)$ in the sample arm and reflected electric field $E_R(k)$ in the reference arm are given by

\begin{equation}
    E_S(k) = E_i(k) \cdot \int_{z_0}^{z_0 + \Delta z}{r_S(z) e^{-j 2k z} \dif z}
\end{equation}
and
\begin{equation}
    E_R(k) = E_i(k) \cdot r_R
\end{equation}
where $k$ is the wavenumber, $z$ is the displacement between the reference mirror and sample, $E_i(k)$ is the incident electric field, and $r_R$ is the reflectivity of the reference mirror. Here, we implicitly place the reference mirror at $z = 0$.

The interferometric pattern $I(k)$ formed by a typical FD-OCT (before being measured) could be expressed as,
\begin{align}
    \begin{split}
        I(k) & = |E_S(k) + E_R(k)|^2\\
        & = |E_S(k)|^2 + |E_R(k)|^2 + E_S^*(k) E_R(k) + E_S(k) E_R^*(k)\\
        & = \textrm{autocorrelation} + \textrm{DC} + S(k) \cdot r_R \cdot \int_{-z_0 - \Delta z}^{z_0 + \Delta z}{f(z) e^{-j 2k z} \dif z} 
    \end{split}
    \label{Equation3}
\end{align}
where $S(k) = E_i^*(k) E_i(k)$ is the power emission spectrum of the light source. Generally speaking, the autocorrelation term could be omitted when the FD-OCT system is operated in shot-noise limited regime and the DC term could be filtered out. Therefore, Eq. (\ref{Equation3}) could be rewritten as,
\begin{equation}
    I(k) = S(k) \cdot r_R \cdot \int_{-z_0 - \Delta z}^{z_0 + \Delta z}{f(z) e^{-j 2 k z} \dif z}
    \label{Eq1}
\end{equation}
where
\begin{align}
    \begin{split}
        f(z) = 
        \left\{
        \begin{matrix}
            & r_S(z),  & z \in [z_0, z_0 + \Delta z] \\
            & r_S(-z),  & z \in [-z_0 - \Delta z, -z_0]\\
            & 0,         & \textrm{otherwise}
        \end{matrix}
        \right.
    \end{split}
\end{align}

The newly constructed object function $f(z)$ is just the original sample profile plus its mirror. Clearly, the spectral interferogram $I(k)$ and object function $f(z)$ forms a Fourier transform pair.

Once the interferogram is measured, $I(k)$ will be digitized, i.e. both truncated \emph{and} sampled. Without losing generality, we assume the signal is uniformly sampled in the $k$ domain by $M$ times. The resultant discrete signal could then be written as,
\begin{align}
    \begin{split}
        I[m] & = I(k_m) + N(k_m)\\
        & = S(k_m) \cdot \int_{-z_0-\Delta z}^{z_0 + \Delta z}{f(z) e^{-j2 k_m z} \dif z} + N(k_m), \quad m = 0, \ldots, M - 1,\\
        & = 2 S(k_m) \cdot \int_{z_0}^{z_0 + \Delta z}{r_S(z) \cos{(2 k_m z)} \dif z} + N(k_m), \quad m = 0, \ldots, M - 1,
    \end{split}
    \label{Eq2}
\end{align}
where $k_m = k_0 + m \delta_k$, $k_0$ is the starting wavenumber for the spectral measurement, $\delta_k$ is the sampling interval in the $k$ domain, and $m $ is the corresponding index. And constant scalars, $\rho$ and $r_R$, are ignored in our analysis.

It is now intriguing for us to quickly evaluate the effect of digitization based on Eq. (\ref{Eq2}): the analog signal $I(k)$ is not only sampled at an interval of $\delta_k$ but also truncated to a shorter range $[k_0, k_{M-1}]$. According to the property of Fourier transform, the first operation, spectral sampling, is equivalent to create a periodic summation of the original function in the object domain and the periodicity is equal to $1/2\delta_k$. However, the second operation, truncation, would convolve the object function with a complex-valued function, which is surprisingly \emph{shift-variant} in its general form. We will come back to this point in Supplementary Section 2.

\section{Image retrieval in OCT as an inverse problem}

Since the fundamental problem in FD-OCT is to recover the original object profile $r_S(z)$ from its spectral measurement $I[m]$. This could be addressed by solving Eq. (\ref{Eq2}) as an inverse problem, which gives the forward relationship between $I[m]$ and $r_S(z)$.

In order to numerically solve Eq. (\ref{Eq2}), which is an integral equation, we could discretize the integral on its right-hand side (RHS),
\begin{equation}
  \sum_{n=0}^{N-1}{r_S(z_n) \cos{(2k_m z_n)} \cdot \delta_z} , \quad n = 0, \ldots, N - 1,
   \label{Eq3}
\end{equation}
where $z_n = z_0 + n\delta_z$, $\delta_z = \Delta z / N$ is the digitization definition, and $n$ is the corresponding index. It is worth noting that the left Riemann sum presented in Eq. (\ref{Eq3}) only equals to the original integral when the digitization definition $\delta_z$ approaches zero,
\begin{align}
    \begin{split}
    \int_{z_0}^{z_0 + \Delta z} & {r_S(z) \cos{(2 k_m z)} \dif z} \\
    = &\lim_{\delta_z \rightarrow 0}\sum_{n=0}^{N-1}{r_S(z_n)} \cos{(2k_m z_n)} \cdot \delta_z , \quad n = 0, \ldots, N - 1 
    \end{split}
    \label{Eq4}
\end{align}

Unfortunately, it is impossible for us to obtain $\delta_z \rightarrow 0$ (or equivalently $N \rightarrow \infty$); a discretization error $\boldsymbol{\epsilon}_{\textrm{disc}}$ would inevitably occur during the process,
\begin{align}
    \begin{split}
    \int_{z_0}^{z_0 + \Delta z} & {r_S(z) \cos{(2 k_m z)} \dif z} \\
    =  & \sum_{n=0}^{N-1}{r_S(z_n)} \cos{(2k_m z_n)} \cdot \delta_z + \epsilon_{\textrm{disc}}(k_m), \quad n = 0, \ldots, N - 1 
\end{split}
\end{align}

Therefore, a feasible alternative is to reduce $\delta_z$ to a value that is small enough so that $\boldsymbol{\epsilon}_{\textrm{disc}}$ falls below a predefined threshold $\epsilon_0$.

Under this condition, Eq. (\ref{Eq2}) can now be transformed into a set of $M$ summation equations, where $r_S(z_n)$ are the $N$ unknowns we want to solve,
\begin{equation}
\left\{
    \begin{aligned}
        I[0] & = 2 S(k_0) \cdot \sum_{n=0}^{N-1}{r_S(z_n)} \cos{(2k_0 z_n)} \cdot \delta_z + N(k_0) + \epsilon_{\textrm{disc}}(k_0)\\
        I[1] & = 2 S(k_1) \cdot \sum_{n=0}^{N-1}{r_S(z_n)} \cos{(2k_1 z_n)} \cdot \delta_z + N(k_1) + \epsilon_{\textrm{disc}}(k_1)\\
        \vdots\\
        I[M-1] & = 2 S(k_{M-1}) \cdot \sum_{n=0}^{N-1}{r_S(z_n)} \cos{(2k_{M-1} z_n)} \cdot \delta_z + N(k_{M-1}) + \epsilon_{\textrm{disc}}(k_{M-1})
    \end{aligned}
\right.
\label{Eq5}    
\end{equation}

Eq. (\ref{Eq5}) could further be re-arranged into a matrix form as given in Eq. (2) in the main text,

\begin{equation}
    \boldsymbol{i} = \delta_z \cdot (\boldsymbol{s} \odot \mathbf{C} \boldsymbol{r}) + \boldsymbol{n}
    \label{matrix}
\end{equation}
where the discretization error $\boldsymbol{\epsilon}_\textrm{disc}$ is absorbed into the new noise term $\boldsymbol{n}$,

\begin{equation}
    \boldsymbol{n}=\{N(k_0), N(k_1), ..., N(k_{M-1}))\}^\intercal + \boldsymbol{\epsilon}_{\textrm{disc}}
\end{equation}

It is now apparent that the image reconstruction of FD-OCT could be categorized as an inverse problem \cite{bertero1998introduction}: the aim is to restore an $N$-dimensional object $\boldsymbol{r}$ from its $M$-dimensional measurement $\boldsymbol{i}$ by solving a set of linear equations stated in Eq. (\ref{matrix}). For example, a solution to this problem could be found via $\ell2$-norm minimization,
\begin{equation}
    \argmin_{\hat{\boldsymbol{r}}} \lVert \boldsymbol{i} -  \frac{\Delta z}{N} \cdot (\boldsymbol{s} \odot\mathbf{C} \hat{\boldsymbol{r}}) \rVert_2^2
    \label{Equation3}
\end{equation}
where $\hat{\boldsymbol{r}}$ is the reconstructed object discretized on a grid that is \emph{arbitrarily} defined by the user.  

\section{The essence of IDFT-based reconstruction}
Conventionally, the reconstruction is performed by directly applying an IDFT on the measured spectrum $I[m]$. Mathematically, it is equivalent to left multiplying Eq. (\ref{matrix}) by an IDFT matrix

\begin{equation*}
    \mathscr{F}^{-1}=
    \begin{bmatrix}
    \exp{(j2z'_0 k_0)} & \exp{(j2z'_0 k_1)} & \dots & \exp{(j2z'_0 k_{M-1})}\\
    \exp{(j2z'_1 k_0)} & \exp{(j2z'_1 k_1)} & \dots & \exp{(j2z'_1 k_{M-1})}\\
    \vdots&\vdots&\ddots&\vdots\\
    \exp{(j2z'_{M-1} k_0)} & \exp{(j2z'_{M-1} k_1)} & \dots & \exp{(j2z'_{M-1} k_{M-1})}\\
   \end{bmatrix}
   \label{equation3}
\end{equation*}
where $z'_m = z_0 + m \delta_{z'}$ defines the grid in the reconstruction domain, $\delta_{z'} = 1 / \Delta k$ is the grid spacing, and $\Delta z' = M \delta_{z'}$ is the grid length (imaging depth). The obtained reconstruction $\hat{\boldsymbol{r}}_{\textrm{IDFT}}$ could be given by,
\begin{align}
    \begin{split}
    \hat{\boldsymbol{r}}_{\textrm{IDFT}}& = \mathscr{F}^{-1}\boldsymbol{i}\\
    & = \frac{\Delta z}{N} \cdot \mathscr{F}^{-1} (\boldsymbol{s} \odot \mathbf{C} \boldsymbol{r})\\
    & = \frac{\Delta z}{N} \cdot ( \mathscr{F}^{-1}\boldsymbol{s}
    \otimes \mathscr{F}^{-1}\mathbf{C}\boldsymbol{r})
    \end{split} 
    \label{equation4}
\end{align}
where $\otimes$ denotes convolution and we applied convolution theorem in the last step.

For an ideal reconstruction, $\hat{\boldsymbol{r}}_{\textrm{IDFT}}$ should be identical to the object $\boldsymbol{r}$ to enable a perfect imaging. However, the direct IDFT reconstruction would inevitably cause some distortions as shown in Eq. (\ref{equation4}): the object $\boldsymbol{r}$ is first transformed by $\mathscr{F}^{-1}\mathbf{C}$ and then convolved with $\mathscr{F}^{-1}\boldsymbol{s}$.

\subsection{The impact of optical transfer function}

The vector $\boldsymbol{s}$ functions effectively as an optical transfer function (OTF). Its IDFT, $\mathscr{F}^{-1}\boldsymbol{s}$, gives the axial point spread function (PSF) of the system; the full-width-half-maximum (FWHM) of the magnitude of the PSF $h(z_m) = |\mathscr{F}^{-1}\boldsymbol{s}|$ is conventionally defined as the axial resolution of the FD-OCT system based on Rayleigh criterion.

The impact of the OTF $\boldsymbol{s}$ on the reconstructed object $\hat{\boldsymbol{r}}_\textrm{IDFT}$ could be understood from two perspectives. First of all, it is band-pass in the $k-$domain, since all the spectral components of the original object $\boldsymbol{r}$ residing outside the support $[S(k_0), S(k_{M - 1})]$ will be implicitly eliminated. This process is previously believed to be irreversible and physically limiting the axial resolution of FD-OCT in conventional theory \cite{RN127}. In other words, we could achieve a higher axial resolution only by increasing the bandwidth of $\boldsymbol{s}$. Secondly, the band-passed portion of $\boldsymbol{r}$ is further modulated by the envelope of $\boldsymbol{s}$. This effect, on the other hand, is often regarded as reversible. Techniques including spectral shaping \cite{Tripathi:02} or deconvolution are proposed to alleviate it.

\subsection{The impact of grid discretization}

As to $\mathscr{F}^{-1}\mathbf{C}$, there is no prior literature discussing its impact to our best knowledge. To analytically study its behaviour, we can explicitly write out its expression \cite{Ling:17},
\begin{align}
    \begin{split}
        & G_{m, n} = {[\mathscr{F}^{-1}\mathbf{C}]}_{m, n} = g_{z_n}(z'_m) = \\
        & \left\{
        \begin{matrix}
            & M,  & z'_m = \pm z_n,\\[4mm]
            &
            \begin{aligned}
            &  \exp{[-j2(k_0 + \frac{M-1}{2}\delta_{k})z_n]}\frac{\sin{[M(\delta_k z_n - m \pi /M)]}}{\sin{(\delta_k z_n - m \pi /M)}}\\[3mm]
            + & \exp{[j2(k_0 + \frac{M-1}{2}\delta_{k})z_n]}\frac{\sin{[M (\delta_k z_n+ m \pi /M)]}}{\sin{(\delta_k z_n + m \pi /M)}},
            \end{aligned}
            & \textrm{otherwise.}
        \end{matrix}
        \right.
    \end{split}
\end{align}
which essentially defines a \emph{shift-variant} mapping from $\mathbb{R}^{\textrm{N} \times 1}$ (object domain) to $\mathbb{C}^{\textrm{M} \times 1}$ (reconstruction domain): for example, Kronecker Delta functions positioned at different spatial locations $z_n$ in the object domain would give different responses $g_{z_n}(z'_m)$ in the reconstruction domain after being transformed by $\mathscr{F}^{-1}\mathbf{C}$. It is also interesting to notice that $g_{z_n}(z'_m)$ consists of two mirrored parts: just as we often observed in OCT reconstructions. In this note, we will confine our discussion on the one in the positive half plane of $z$,

\begin{equation}
    g_{z_n,+}(z'_m) = \exp{[-j2(k_0 + \frac{M-1}{2}\delta_{k})z_n]}\frac{\sin{[M(\delta_k z_n - m \pi /M)]}}{\sin{(\delta_k z_n - m \pi /M)}}
    \label{eqn:general}
\end{equation}

The first term, $\exp{\{-j2(k_0 + (M-1)/2 \delta_{k})z_n\}}$, is a complex exponential function, which determines the phase of the reconstructed object $\hat{\boldsymbol{r}}_{\textrm{IDFT}}$. Since its phase is linearly modulated by the $z_n$, one can localize the original object at a precision even finer than a fraction of the incident wavelength. This essentially manifests the foundation of the so-called phase-sensitive or phase-resolved OCT techniques \cite{Choma:05}. However, this type of techniques are only suitable for isolated objects, and a phenomenon known as signal competition would emerge if multiple samples are placed closely \cite{Lin}.

The second term, $\sin{[M (\delta_k z_n - m \pi /M)]}/\sin{(\delta_k z_n - m \pi /M)}$, is the Dirichlet kernel, which is the sampling (or interpolation) kernel for the IDFT. Both terms are \emph{shift-variant} in their general forms.

\subsubsection{Case I: matched grids}
Let's now consider the first case that the sampling grid in the object domain exactly matches that in the reconstruction domain,
\begin{equation}
    \left \{ 
    \begin{aligned}
        \delta_z & = \delta_{z'}\\
        M & = N
    \end{aligned}
    \right.
\end{equation}

Therefore, the $g_{z_n, +}(z'_m)$ will become a sinc function for a given $z_n$, and matrix $G$ would degenerate to an Identity matrix. In this case, the reconstructed $\hat{\boldsymbol{r}}_{\textrm{IDFT}}$ will only be degraded by the axial PSF.

\subsubsection{Case II: mismatched grids cross the domains}

Unfortunately, it is extremely unlikely, if not possible, to encounter the scenario depicted in case I: the grids will hardly be matched cross the domains in real-world scenarios, since the granularity of the original object function is generally much finer than the coarse grid defined by the
IDFT. Therefore, we need to tackle the general form of $g_{z_n, +}(z'_m)$ as that in Eq. (\ref{eqn:general}).

\begin{figure}
    \centering
    \includegraphics[width = \textwidth]{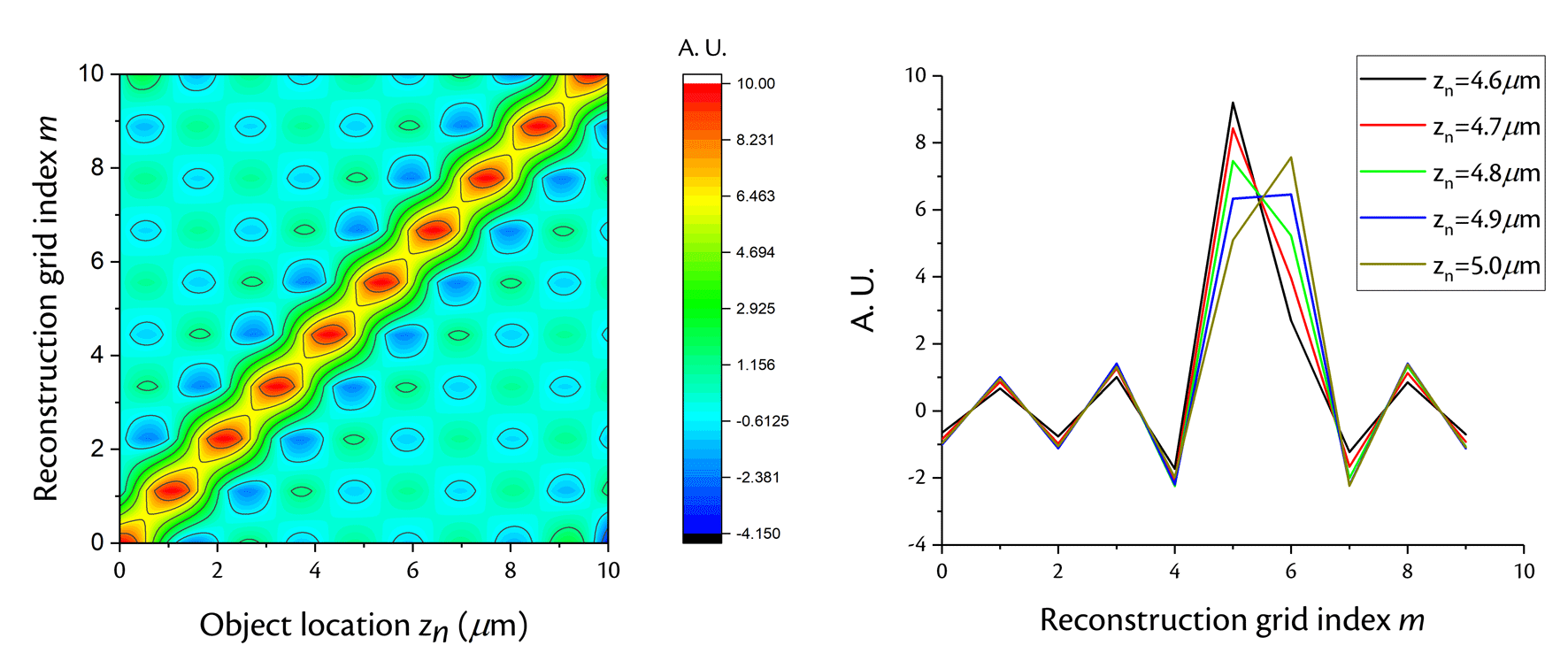}
    \caption{The IDFT-based reconstruction is \emph{shift-variant}. (a) The Dirichlet kernel manifests different shape for different input locations. (b) We plot the reconstruction response against input object location as a two dimensional contour image.}
    \label{fig:my_label}
\end{figure}

Recall the Dirichlet kernel, $\sin{[M (\delta_k z_n - m \pi /M)]}/\sin{(\delta_k z_n - m \pi /M)}$, is a \emph{shift-variant} function: it would change its shape upon different $z_n$. To illustrate this point, we set $z_n$ to be 4.6 $\mu$m, 4.7 $\mu$m, 4.8 $\mu$m, 4.9 $\mu$m, and 5.0 $\mu$m, and plot the corresponding Dirichlet kernel in Fig. \ref{fig:my_label}a. $\delta_k$ is assumed 2094.1 m$^{-1}$. It is obvious that the system is \emph{shift-variant} as we expected. We also plot the Dirichlet kernel as a color-coded contour image as shown in Fig. \ref{fig:my_label}b for reference.

It is quite clear now that convolving the object $\boldsymbol{r}$ with $g_{z_n, +}(z'_m)$ would impair the imaging quality of the system in a special way: since $g_{z_n, +}(z'_m)$ is \emph{shift-variant}, we could not use conventional tools such as deconvolution to resolve it. This also partially explained why the deconvolution is not that popular in OCT communities.

\subsection{Summary}

To summarize, the IDFT-based technique provides us with a simple and satisfactory reconstruction result. It is now quite clear that the conventional IDFT-based reconstruction solves the inverse problem as stated in Eq. (\ref{matrix}) in a special case: it offers a least-squares solution, if we ignore the modulation of the emission spectrum (let $\boldsymbol{s}$ be an all-one vector) and if the discretization grid in the object domain is identical to that of the reconstruction domain. 

However, in general, the axial resolution of IDFT-based reconstruction is limited and the accuracy of the reconstruction is degraded by a \emph{shift-variant} impulse response. It could be understood as follows.





First of all, a large portion of the equations in Eq. (\ref{matrix}) are ``wasted'' in the IDFT-based reconstruction scheme. Considering that it is rather common to have a much larger imaging depth $\Delta z'$ than the object's support $\Delta z$ (due to the shallow penetration depth of OCT), a good number of unknown variables $\hat{r}[m]$ are doomed to be zeros as illustrated in main text Fig. 1.

Secondly, the IDFT-based reconstruction is purely mathematical with no prior knowledge imposed. Intuitively, the proposed optimization-based technique should in theory outperform the IDFT-based technique due to the integration of physical insights.

\section{The theoretical limit of the proposed technique: a numerical study}

One important question that we have not addressed in the main text is what the theoretical axial limit of the proposed technique would be. Although we believe a complete discussion on this topic is well out of the scope of this study, we would still like to present a preliminary numerical study to press this issue.

\subsection{Qualitative study: noise-free reconstruction}

\begin{figure}[htbp]
        \includegraphics[width=\textwidth]{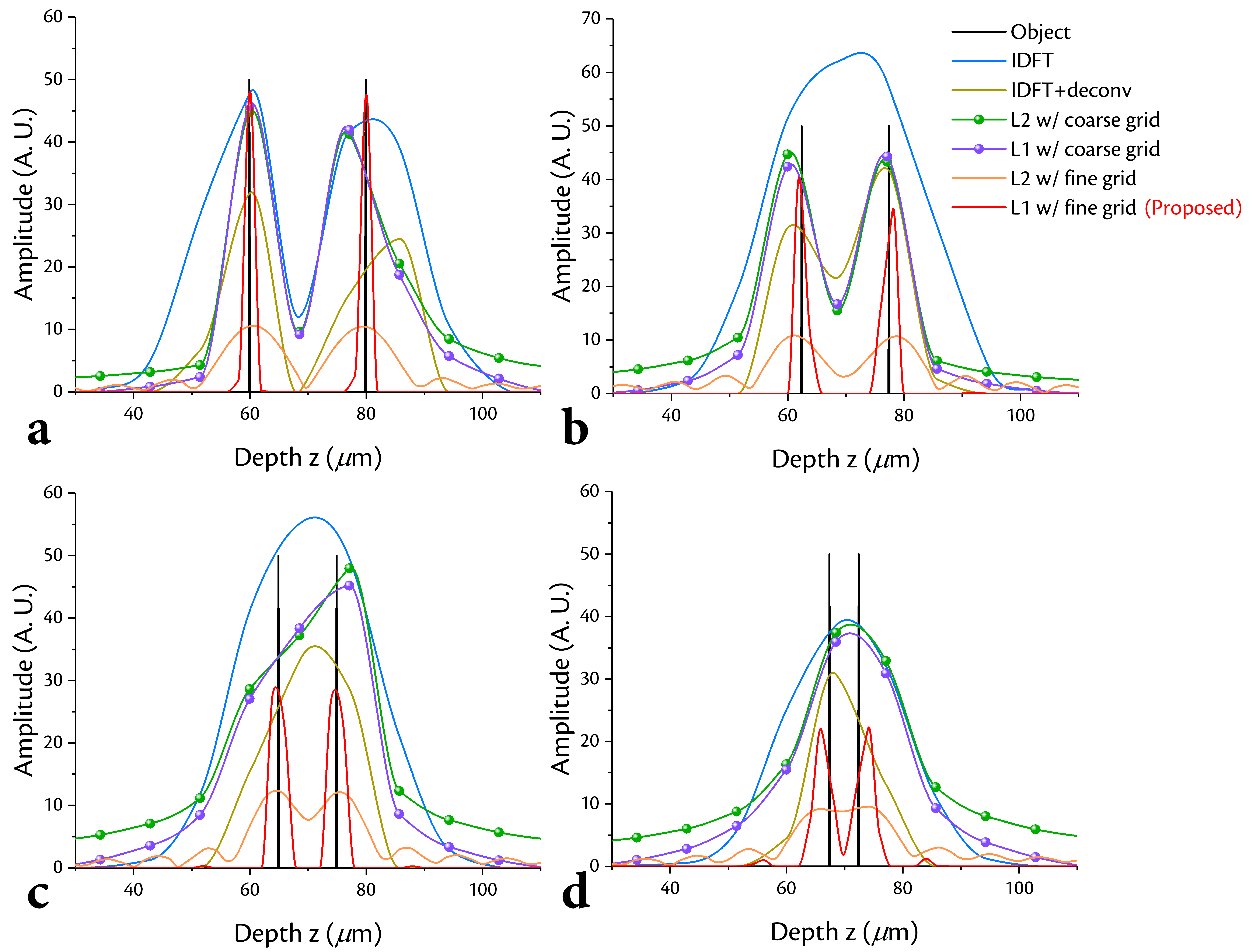}
        \caption{Simulation of six different reconstruction techniques for FD-OCT. Two Kronecker Delta functions, the object, are separated by (a) 20 $\mu$m, (b) 15 $\mu$m, (c) 10 $\mu$m, and (d) 5 $\mu$m, respectively. The theoretical axial resolution of the simulated system is $\sim$19$\mu$m. The proposed $\ell 1$-norm minimization with fine grid shows the best performance.}
    \label{fig:spikes}
\end{figure}

We first assumed that there is no noise in the system, and we tried to depict a visual picture to illustrate the resolving capability of different techniques. We used two closely located Kronecker Delta functions as the target and numerically synthesized their spectral interferograms. Various techniques including including plain IDFT, plain IDFT followed by Lucy-Richardson deconvolution \cite{Liu:09}, $\ell1$- and $\ell2$-norm minimization with both coarse \cite{Mousavi:16} and fine grid were used to reconstruct the object. Here, the coarse grid is same as that defined by the Fourier relationship, while the fine grid's step is $\sim$8 times smaller. The reconstructed results are plotted in Fig. \ref{fig:spikes} for different separations. Considering that the theoretical axial resolution of the system is $\sim$19 $\mu$m (Supplementary Section 4.3), it is expected that the conventional IDFT-based reconstruction could only resolve the spikes when they are separated by 20 $\mu$m as shown in Fig. \ref{fig:spikes}a. The resolving capability of the system is slightly improved when deconvolution algorithm is applied afterwards: the 15 $\mu$m’s separation could be discriminated as illustrated in Fig. \ref{fig:spikes}b. Neither $\ell 1$- nor $\ell 2$-norm minimization with coarse grid obtains better performance than that of the deconvolution. However, using $\ell 2$-norm minimization in a finer grid succeeds in resolving spikes that are only 10 $\mu$m apart, and using $\ell 1$-norm minimization in a finer grid further push the limit to 5 $\mu$m. Resolution beyond theoretical prediction is achieved in both cases, and they are depicted in Fig. \ref{fig:spikes}c and \ref{fig:spikes}d, respectively.

\subsection{Quantitative study: noisy reconstruction}

We then carried out a quantitative study to evaluate the impact of various parameters including Lagrange multiplier $\lambda$, grid step $\delta_z$, and signal-to-noise ratio (SNR) on the reconstruction resolution. Specifically, Monte Carlo method was used to simulate the system's behaviour in noisy settings. We applied the same definition for axial resolution as described in main text's Methods which is much more restrict than the visual inspection: a minimal separation (i.e. axial resolution) is logged if the two peaks are separated based on Rayleigh criterion \emph{and} the relative error between the measurement and the ground truth is less than 20\%.

Although we still used two Kronecker Delta functions as the object, their depth locations were randomly generated in this experiment from trial to trial. 

\begin{figure}[h]
\centering
\includegraphics[width = 0.8 \textwidth]{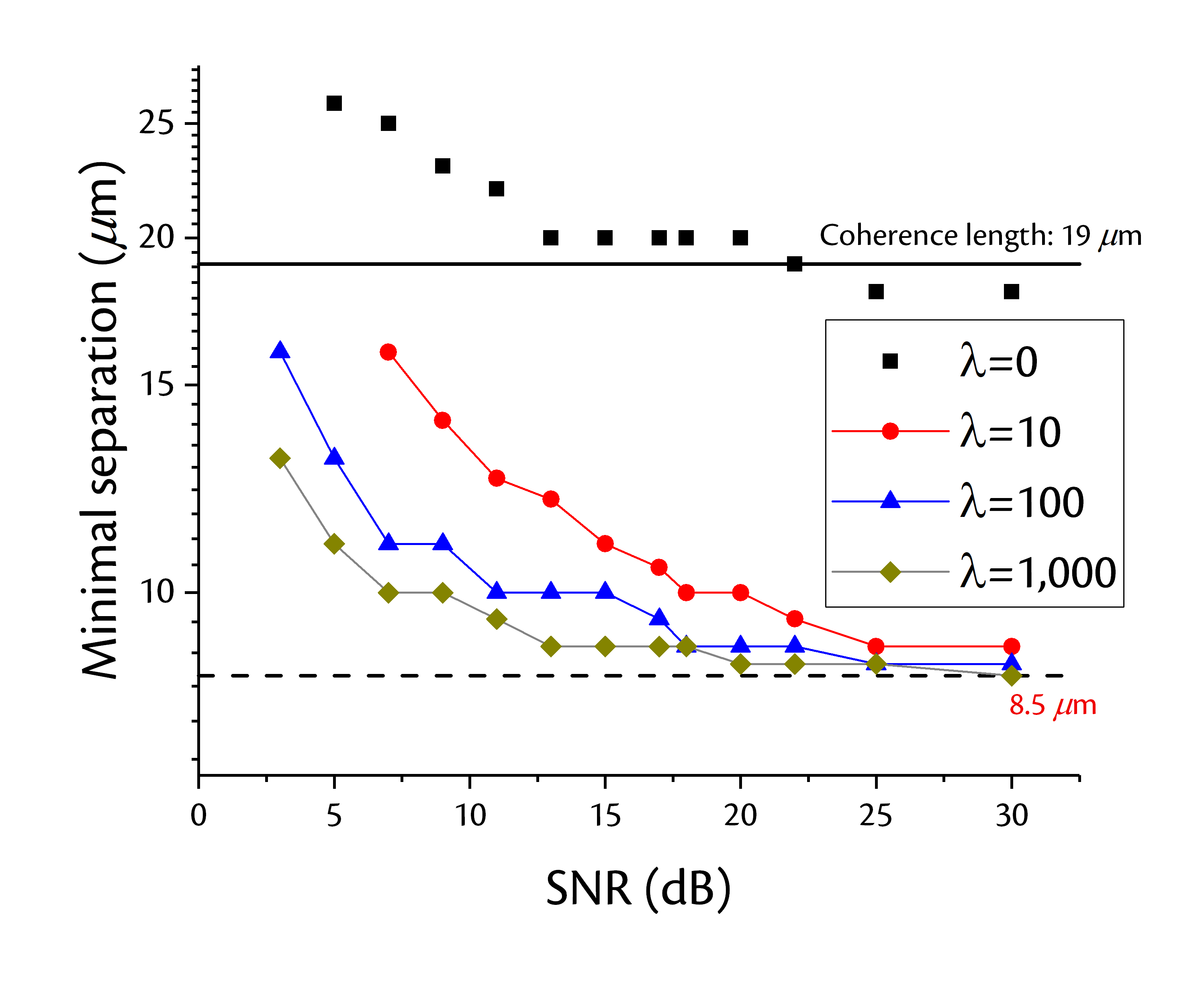}
\caption{The minimal separations under different values of $\lambda$ by using the proposed technique are plotted against SNR. A solid horizontal line gives the theoretical axial resolution of the system, which is $\sim$19 $\mu$m. The dashed horizontal line gives a hypothetical asymptotic value for the proposed technique with increasing $\lambda$ and increasing SNR. It is clear that the proposed technique consistently beat the theoretical limit even at a very low SNR. It is worth noting that the $\lambda = 0$ case is corresponding to the $\ell2$-minimization scheme \cite{Mousavi:16}.}
\label{fig:lambda}
\end{figure}

The impact of choosing different $\lambda$ is shown in Fig. \ref{fig:lambda}, in which the grid step was fixed at 1 $\mu$m. It is clear that the minimal separation turns smaller when SNR or $\lambda$ increases, and it asymptotically converges to $\sim$8.5 $\mu$m. Besides, a larger Lagrange multiplier $\lambda$ could also performs better in low SNR regime, which suggests a better denoising capability. In contrast, the minimal separation detected by using $\ell2$-minimization ($\lambda = 0$) is not significantly better than the theoretical prediction. We believe this phenomenon is due to the lack of regularization when solving Eq. (\ref{matrix}).

\begin{figure}[h]
\centering
\includegraphics[width = 0.8\textwidth]{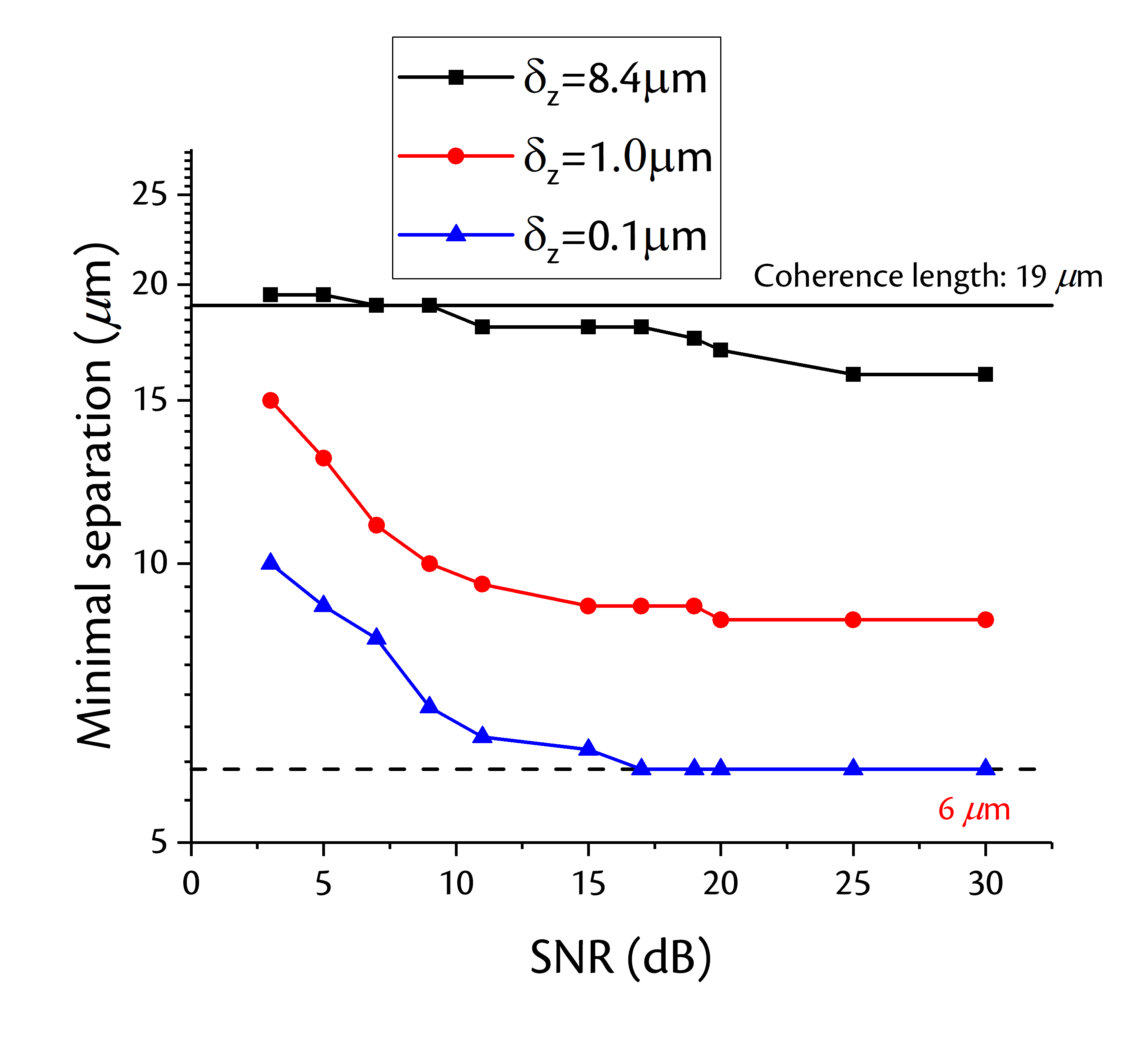}
\caption{The minimal separations under different values of $\delta_z$ by using the proposed technique are plotted against SNR. All results were obtained by setting $\lambda$ to be 500. A solid horizontal line gives the theoretical axial resolution of the system, which is $\sim$19 $\mu$m. The dashed horizontal line gives the best resolution achieved by the the proposed technique with increasing SNR.}
\label{fig:dz}
\end{figure}

We then turned to study the effect of grid step $\delta_z$. We fixed the $\lambda$ to be 500 and set the grid step to be 0.1 $\mu$m, 1 $\mu$m, and 8.4 $\mu$m, respectively. It is worth noting that we choose $\delta_z=8.4$ $\mu$m because it is the same as the pre-defined grid step by Fourier relationship. The result is plotted in Fig. \ref{fig:dz}. As expected, by using a finer grid step, we could obtain higher resolution, which could be ascribed to smaller approximation errors. Specifically, the best performance is achieved when the finest grid (0.1 $\mu$m) is used: the resultant minimal separation of about 6 $\mu$m is almost 3 times better than that of the default grid. Unfortunately, we failed to further our study by using a even smaller $\delta_z$ due to the hefty computational cost. 

\subsection{Numerical simulation set-up}
For the simulation, we assume the power spectral density $S(k)$ have a Gaussian distribution, the central wavelength $\bar{\lambda}$ is 1310 nm, and the FWHM is 40 nm. This gives a coherence length about 19 $\mu$m. The spectral interferogram is sampled by $M = 1,024$ times within a range of 100 nm (from 1260 nm to 1360 nm).

The discretization interval in the object domain was set to be an extremely small value (100 nm here) to minimize the approximation error. The support size of the object was set to 1 mm. We numerically computed the spectral measurement $I[m]$ as the input. For noisy cases, an additive Gaussian noise is added.

For reconstruction, the size of the coarse grid was equated to the reconstruction definition in the IDFT case ($\sim$8.4 $\mu$m), while the size of the fine grid to be much smaller ($\delta_z = 1$ $\mu$m). We used the built-in function \verb!IFFT! and \verb!deconvlucy! in OCTAVE to implement the IDFT and Lucy-Richardson algorithm, respectively. The $\ell1$- and $\ell2$-norm minimization are both implemented by the ADMM-based algorithm described in the Methods; $\lambda$ was set to a non-zero value for $\ell1$-norm minimization and 0 for $\ell2$. All the computations were performed on the SJTU high-performance computing facilities.

All the data points presented in Fig. \ref{fig:lambda} and \ref{fig:dz} are averaged value from 10 experiments.

\bibliographystyle{naturemag}
\bibliography{ref}